\documentclass[aps,pra,twocolumn,amsmath,amssymb,showpacs,superscriptaddress]{revtex4}

\newcommand{\bra}[1]{\langle#1|}
\newcommand{\ket}[1]{|#1\rangle}
\usepackage[dvips]{graphicx}
\usepackage{mathrsfs}

\begin{document}

\title{Practical limitations in optical entanglement purification}

\author{Peter P. Rohde}
\email[]{rohde@physics.uq.edu.au}
\homepage{http://www.physics.uq.edu.au/people/rohde/}
\affiliation{Centre for Quantum Computer Technology, Department of Physics\\ University of Queensland, Brisbane, QLD 4072, Australia}

\author{Timothy C. Ralph}
\affiliation{Centre for Quantum Computer Technology, Department of Physics\\ University of Queensland, Brisbane, QLD 4072, Australia}

\author{William J. Munro}
\affiliation{Hewlett-Packard Laboratories, Filton Road, Stoke Gifford, Bristol BS34 8QZ, UK}

\date{\today}

\begin{abstract}
Entanglement purification protocols play an important role in the distribution of entangled systems, which is necessary for various quantum information processing applications. We consider the effects of photo-detector efficiency and bandwidth, channel loss and mode-mismatch on the operation of an optical entanglement purification protocol. We derive necessary detector and mode-matching requirements to facilitate practical operation of such a scheme, without having to resort to destructive coincidence type demonstrations.
\end{abstract}


\maketitle

Entanglement between distributed quantum systems is of fundamental importance to the future implementation of various quantum information processing devices and applications, including quantum teleportation, quantum cryptography and distributed quantum computing. Such applications typically place strict requirements on the required fidelity of the entangled systems. For this reason entanglement purification schemes \cite{bib:Bennett96,bib:Deutsch96,bib:Bose99,bib:Duan00,bib:Pan01} are of practical interest, since they allow us to purify a desired entangled state with higher fidelity from multiple copies which have been subject to noise and are of lesser fidelity.

Here we consider a recent proposal for optical entanglement purification \cite{bib:Pan01} which has been subject to in-principle experimental demonstration \cite{bib:Pan03}. In theory, the proposal of Ref.~\cite{bib:Pan01} requires unit efficiency photo-detectors with infinite bandwidth and assumes a loss-less channel between successive rounds. In practice, none of these assumptions are accurate. This has the effect of introducing ambiguity into the number of photons responsible for a given detection event. For example, in the presence of detector inefficiency a two photon event may be perceived as a single photon event. For this reason the operation of such protocols is highly susceptible to detector inefficiency and photon loss. This necessitates operating present day experiments, such as that reported in Ref.~\cite{bib:Pan03}, in coincidence. This eliminates photon number ambiguity by post-selecting away undesired detection events. While this is satisfactory for in-principle demonstrations and some applications, there are applications in which post-selection on detecting the correct number of photons is not possible. For example, frequently touted applications for entangled photons include entanglement distribution between distant atomic systems and solid-state quantum memory \cite{bib:Chaneliere05,bib:Matsukevich05,bib:Matsukevich05b}. Here photons couple into trapped atoms, inducing particular atomic transitions, thereby mapping photonic states to atomic states. Because the photons are absorbed by the system, it is not possible to post-select on the desired photon-number.

In this paper we examine the effects of channel loss, photo-detector efficiency, bandwidth, and mode-mismatch on the operation of this entanglement purification protocol. We derive the experimental requirements to allow such a scheme to operate effectively in a non-coincidence environment. Our results indicate that experimental requirements in this context are very demanding and will require significant effort.

\textbf{Optical entanglement purification:} The entanglement purification protocol described in Ref.~\cite{bib:Pan01} relies on resource states which are assumed to be mixtures of the form
\begin{equation} \label{eq:resource_state}
\hat\rho_\mathrm{in}=F\ket{\Phi^+}\bra{\Phi^+}+(1-F)\ket{\Psi^+}\bra{\Psi^+}
\end{equation}
$\ket{\Phi^\pm}=\frac{1}{\sqrt{2}}(\ket{H}_A\ket{H}_B\pm\ket{V}_A\ket{V}_B)$ and
$\ket{\Psi^\pm}=\frac{1}{\sqrt{2}}(\ket{H}_A\ket{V}_B\pm\ket{V}_A\ket{H}_B)$ are the usual Bell states, where $\ket{H}$ and $\ket{V}$ denote the horizontally and vertically polarized single photon states, and $A$ and $B$ the two parties. $F$ is the fidelity of the state with respect to the desired entangled state $\ket{\Phi^+}$, \emph{i.e.} \mbox{$F=\bra{\Phi^+}\hat\rho_\mathrm{in}\ket{\Phi^+}$}. It is evident from Eq.~\ref{eq:resource_state} that the error model assumes a single bit-flip error (the generality of the model will be discussed later).

\begin{figure}[!htb]
\includegraphics[width=\columnwidth]{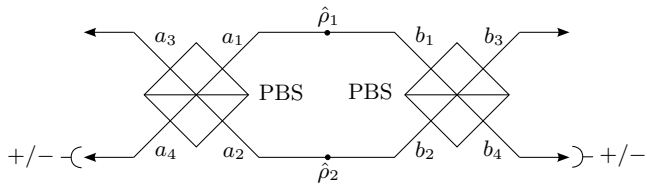}
\caption{Experimental layout of the optical entanglement purification protocol. A single purified state (modes $a_3$ and $b_3$) is generated from two pairs with lower fidelity ($\hat\rho_1$ and $\hat\rho_2$). Parity measurements are implemented using the PBS's followed by post-selection ($+/-$) on modes $a_4$ and $b_4$ in the diagonal/anti-diagonal polarization basis.} \label{fig:exp_layout}
\end{figure}

The protocol employs two copies of the resource state. The joint input state can therefore be expressed
\begin{eqnarray}
\hat\rho_\mathrm{joint}&=&F^2\ket{\Phi^+}_1\ket{\Phi^+}_2\bra{\Phi^+}_1\bra{\Phi^+}_2\nonumber\\
&+&F(1-F)\ket{\Phi^+}_1\ket{\Psi^+}_2\bra{\Phi^+}_1\bra{\Psi^+}_2\nonumber\\
&+&F(1-F)\ket{\Psi^+}_1\ket{\Phi^+}_2\bra{\Psi^+}_1\bra{\Phi^+}_2\nonumber\\
&+&(1-F)^2\ket{\Phi^+}_1\ket{\Phi^+}_2\bra{\Phi^+}_1\bra{\Phi^+}_2
\end{eqnarray}
Non-deterministic parity measurements are performed between the corresponding modes of the two copies of $\hat\rho_\mathrm{in}$ using polarizing beamsplitters (PBS's) and post-selection, as shown in Fig.~\ref{fig:exp_layout}. With suitable local operations (dependent on the measurement outcomes), this projects the total state into the subspace where both photons received by a given party have the same polarization. This effectively eliminates $\ket{\Phi^+}_1\ket{\Psi^+}_2$ and $\ket{\Psi^+}_1\ket{\Phi^+}_2$ terms from the joint state, thereby increasing the overall fidelity. This improvement in fidelity comes at the expense of success probability, which is 25\% (the success probability of each Bell measurement is 50\%).

\begin{figure}[!htb]
\includegraphics[width=0.8\columnwidth]{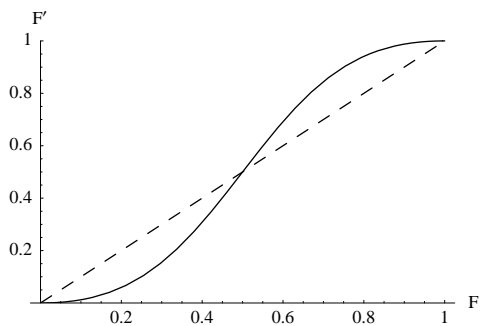}
\caption{$F'$ against $F$ for the entanglement purification protocol (solid line), and $F'=F$ (dashed line). The protocol improves state fidelity in the regime where $F>\frac{1}{2}$.} \label{fig:F_ideal}
\end{figure}

The output state, between modes $a_3$ and $b_3$, can be expressed
\begin{equation}
\hat\rho_\mathrm{out}=F'\ket{\Phi^+}\bra{\Phi^+}+\left(1-F'\right)\ket{\Psi^+}\bra{\Psi^+}
\end{equation}
where $F'$ is the fidelity of the output state and is related to the input state fidelity by
\begin{equation} \label{eq:fidelity}
F'=\frac{F^2}{F^2+(1-F)^2}
\end{equation}
For the protocol to improve the fidelity of the output state it is required that $F'>F$, which is satisfied when \mbox{$F>\frac{1}{2}$}. Thus, provided the fidelity of the resource states being employed is above 50\% the protocol can, in principle, purify the states. Fig.~\ref{fig:F_ideal} shows the relationship between $F'$ and $F$.

The scheme described allows for entanglement purification assuming a single bit-flip error model. However bit-flip errors can be transformed into phase-flip errors via Hadamard gates. Thus, through two consecutive applications of the purification scheme, once in a rotated basis, the scheme can purify full depolarizing noise.

The scheme described can be cascaded to iteratively increase the fidelity of resource states. In this case, the number of resource states required scales as $4^n$, where $n$ is the number of iterations applied and we assume depolarizing noise.

\textbf{Channel loss and photo-detector efficiency effects:} We begin by considering the effects of channel loss and detector inefficiency on the operation of the protocol. First we introduce a more general form for the resource state, which contains additional terms where photons have been lost.
\begin{eqnarray} \label{eq:general_form}
\hat\rho_\mathrm{in}&=&P_2\left[F\ket{\Phi^+}\bra{\Phi^+}+(1-F)\ket{\Psi^+}\bra{\Psi^+}\right]\nonumber\\
&+&P_1\ket{\psi_\mathrm{loss}}\bra{\psi_\mathrm{loss}}+P_0\ket{0}\bra{0}
\end{eqnarray}
where $P_n$ is the probability that the state contains $n$ photons, $\ket{\psi_\mathrm{loss}}$ collectively represents all states containing a single photon (\emph{i.e.} where a single photon has been lost), and $\ket{0}$ is the vacuum state, where both photons have been lost.

We now consider the propagation of such resource states through the entanglement purification protocol. We assume the two resource states are always identical. In cases where no photons have been lost the protocol will operate as previously and with a success probability of $1/4$. Where one of the resource states contains two photons and the other a single photon, the output state will be deficient by one photon, with a success probability of $1/2$. Where both resource states contain a single photon, the output state will contain no photons, with a success probability of $1/8$. Where one resource state contains two photons and the other no photons, the output will also contain no photons, with a success probability of $1/2$. Cases where one resource state contains no photons and the other no or exactly one photon will never succeed.

Based on these observations the output state will be of the form
\begin{eqnarray}
\hat\rho_\mathrm{out}&=&P_2'\left[F'\ket{\Phi^+}\bra{\Phi^+}+(1-F')\ket{\Psi^+}\bra{\Psi^+}\right]\nonumber\\
&+&P_1'\ket{\psi_\mathrm{loss}}\bra{\psi_\mathrm{loss}}+P_0'\ket{0}\bra{0}
\end{eqnarray}
where
\begin{equation} \label{eq:probs}
P_2'=\frac{1}{4}P_2^2,\ P_1'=\frac{1}{2}P_1P_2,\ P_0'=\frac{1}{8}P_1^2+\frac{1}{2}P_2P_0
\end{equation}
and $F'$ is defined as previously.

We now apply a lossy channel to each of the output arms, characterized by intensity-loss $\eta$. This will transform the output probabilities according to
\begin{eqnarray} \label{eq:loss_probs}
P_2''&=&P_2'(1-\eta)^2\nonumber\\
P_1''&=&P_2'2\eta(1-\eta)+P_1'(1-\eta)\nonumber\\
P_0''&=&P_0'+P_1'\eta+P_2'\eta^2
\end{eqnarray}
This is equivalent to utilizing photo-detectors with efficiency $1-\eta$.

Thus, while the presence of loss does not affect the fidelity of the output state (post-selected against states containing the correct number of photons), the probability of the output state containing the correct number of photons is highly loss-dependent. This is expected, since the protocol is designed to protect against Pauli errors, not photon loss.

The term of interest in the output state is $P_2$, corresponding to the probability that the state has the desired two photons. From Eqs.~\ref{eq:probs} and \ref{eq:loss_probs}, following $n$ iterations of the scheme the net probability of the output state containing two photons will be
\begin{equation}
P_2(n)=\left[\frac{1}{4}(1-\eta)^2\right]^{2^n-1}
\end{equation}
using the initial condition $P_2(1)=1$. This corresponds to the scaling of success probability in a post-selected sense. In this context the fidelity of the purified entangled pair is given by $n$-fold iterative application of Eq.~\ref{eq:fidelity}.

\begin{figure}[!htb]
\includegraphics[width=\columnwidth]{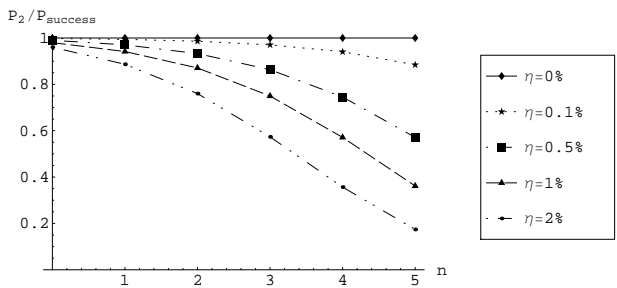}
\caption{Probability of obtaining a two-photon state, normalized against post-selection success probability.} \label{fig:prob_2_photons}
\end{figure}

In a non-post-selected environment this approach to calculating fidelity is not appropriate. For example, if we are writing our purified entangled pair into a solid state quantum memory, in a situation where there is no well defined loss signature, clearly the fidelity of the state in memory will also depend on the probability of having the correct number of photons. Another example is linear optics quantum computation where Bell pairs of high fidelity and efficiency are required \cite{bib:KLM01,bib:RalphHayes05}. In such cases a more relevant question to ask is `what is the probability of having a two-photon state given that the post-selection procedure was successful?'. This is shown in Fig.~\ref{fig:prob_2_photons}.
The effective fidelity of the output state can now be treated as $F\times P_2/P_\mathrm{success}$, where $P_2/P_\mathrm{success}$ is the probability of there being two photons, normalized against success probability. This imposes very stringent loss/efficiency constraints, since $P_2/P_\mathrm{success}$ effectively places an upper bound on the achievable fidelity. For example, assuming three applications of the purification scheme \footnote{Three rounds is of natural interest since this is the number required to perform one round of purification against arbitrary depolarizing noise -- once to protect against bit-flip errors, then again in a rotated basis to protect against phase-flip errors.} and a loss rate of only 1\%, $P_2/P_\mathrm{success}$ is approximately 75\%.

\textbf{Photo-detector spectral bandwidth effects:} We now consider the effect of photo-detector spectral bandwidth. We model this in the same manner as described in Ref.~\cite{bib:RohdeRalph05c}. Specifically, we precede an ideal photo-detector with a beamsplitter with frequency dependent beamsplitter transmissivity $\eta(\omega)$.

The ideal photo-detector is assumed to respond to all spectral components without being able to discriminate between them. The frequency dependent beamsplitter effectively truncates the range of frequencies the detector is able to respond to. Such a photo-detector is described by the measurement projector
\begin{equation}
\hat\Pi_\mathrm{click}=\int_{-\Omega}^{\Omega}\hat{a}^\dag(\omega)\ket{0}\bra{0}\hat{a}(\omega)\,\mathrm{d}\omega=\int_{-\Omega}^{\Omega}\ket{\omega}\bra{\omega}\,\mathrm{d}\omega
\end{equation}
where $\omega$ is frequency, $\hat{a}(\omega)$ is the frequency specific photonic annihilation operator, $\ket{\omega}$ is the single frequency single photon state, and $\Omega$ is the spectral bandwidth of the detector. Here we have assumed $\eta(\omega)$ to be a top-hat function with width $2\Omega$. $\Omega$ is in units of photon bandwidth and we assume Gaussian, transform-limited photons. Thus, our model assumes that the detector can only \emph{see} spectral components in the range $-\Omega$ to $\Omega$ and is unable to discriminate between different spectral components within this window. As $\Omega\to\infty$ the beamsplitter modeling spectral bandwidth becomes completely transmissive for all frequencies and the detector approaches an ideal detector. On the other hand, for finite $\Omega$ some spectral components are traced out. This has the same effect as detector inefficiency, since the detector effectively discards some of the wave-packet. The relationship between the detector's spectral bandwidth and effective efficiency is given by $1-\eta=\mathrm{Erf(\Omega/\sqrt{2})}$. Using this relationship the previous results for photon loss apply.

\textbf{Modeling mode-mismatch:} One of the most significant challenges facing the experimental demonstration of optical quantum information processing applications is mode-mismatch, whereby photon indistinguishability is compromised within a circuit \cite{bib:RohdePryde05}. This can be introduced for a variety of reasons including imperfect alignment of optical components (both spatially and temporally), non-identical photon sources, impure photon sources, time-jitter and wave-packet dispersion. Some degree of mode-mismatch is inevitable in any experimental scenario. We now consider the effects this has on the entanglement purification protocol.

We model mode-mistmatch in the manner described by Ref.~\cite{bib:RohdePryde05}, as temporal displacements between beamsplitter inputs. These displacements can be directly interpreted as degrees of temporal mode-mismatch, or time-jitter in the photon sources. However, it has been shown \cite{bib:RohdePryde05} that such a model for photon distinguishability is sufficient to model arbitrary forms of mode-mismatch or photon distinguishability. During beamsplitter interactions it is only relative displacements which are of significance. Therefore, two displacement parameters ($\tau_1$ and $\tau_2$) are sufficient to model arbitrary mode-matching effects -- one between modes $a_1$ and $a_2$, and the other between $b_1$ and $b_2$.

\begin{figure}[!htb]
\includegraphics[width=\columnwidth]{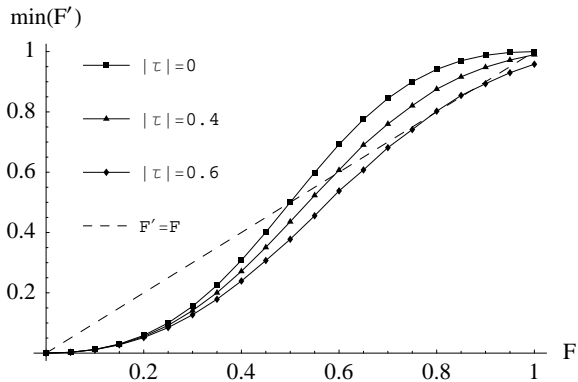}
\caption{Worst case output state fidelity for given upper bounds on the magnitude of temporal mode-mismatch $\tau$ (in units of photon temporal bandwidth).} \label{fig:F_MM}
\end{figure}

For a given magnitude of mode-mismatch ($\tau$) we perform a Monte-Carlo search over $\tau_1$ and $\tau_2$ in the range $-\tau$ to $\tau$. Fig.~\ref{fig:F_MM} shows the relationship between $\mathrm{min}(F')$ and $F$ for various upper bounds on the mode-mismatch magnitude, assuming all else is perfect. It is evident that $\tau=0.6$ (in units of photon temporal bandwidth) represents an upper bound on the degree of allowable mode-mismatch, since for $\tau>0.6$ the output state fidelity is always less that the input state fidelity. This corresponds to a two-photon HOM \cite{bib:HOM87} visibility of $V=0.54$. For $\tau=0.4$ ($V=0.74$), which is within the reach of present day experimental techniques, the scheme is effective over almost the full range of $F$ and the maximum fidelity attainable through successive application of the scheme is almost optimal. However, the fidelity improvement per round is significantly reduced compared to the ideal case. Thus, to achieve comparable fidelity more rounds are needed, which incurs a very substantial penalty in the success probability.

\textbf{Discussion \& conclusion:} We considered the effects of channel loss, photo-detector efficiency and spectral bandwidth as well as mode-mismatch on the operation of an optical entanglement purification protocol. It has been shown that the experimental requirements for such a scheme to operate in a non-coincidence environment are very demanding and represent a significant challenge. The mode-matching requirements are less stringent and within the realm of what is presently possible. Nonetheless, our results indicate that significant care needs to be taken to minimize mode-mismatch.

The formidable loss/efficiency and bandwidth requirements of the discussed scheme come as a direct consequence of it being heralded (depending on the application). This is a common characteristic of heralded quantum information processing circuits and has been observed previously \cite{bib:Lund03,bib:RohdeRalph05}. In stark contrast to this, coincidence type schemes typically benefit from narrowband photo-detection.

In conclusion, many current experimental realizations of photonic quantum information processing protocols operate in coincidence, which mitigates problems associated with photon loss and detector inefficiency. Eventually such schemes will need to be realized in a non-coincidence environment. We have demonstrated that in this context experimental requirements are extremely demanding and even very modest levels of photon loss and detector inefficiency can undermine their operation. This provides insight into where improvements in technology need to be undertaken. Our findings are likely to apply to many other photonic quantum information processing schemes, particularly those that do not operate in coincidence.

This work was supported by the Australian Research Council and the QLD State Government. We thank Sean D. Barrett and Austin Lund for helpful discussions.

\bibliography{paper}

\end{document}